\documentclass[aps,pra,preprint,showpacs,showkeys,superscriptaddress]{revtex4}
\usepackage{pstricks}
\usepackage{epsfig}
\usepackage{graphicx}
\usepackage{amsmath}
\usepackage{amsfonts}
\usepackage{amssymb}
\usepackage{rotating}
\usepackage{url}
\usepackage{hyperref}
\newcommand{\bra}[1]{\langle #1|}
\newcommand{\ket}[1]{|#1\rangle}

\allowdisplaybreaks

\begin{document}
\title{Variational optimization of the 2DM: approaching three-index accuracy using extended cluster constraints.}
\author{Brecht Verstichel}
\email{brecht.verstichel@ugent.be}
\affiliation{ Center for Molecular Modeling, Ghent University,Technologiepark 903, 9052 Zwijnaarde, Belgium}
\author{Ward Poelmans}
\affiliation{ Center for Molecular Modeling, Ghent University,Technologiepark 903, 9052 Zwijnaarde, Belgium}
\author{Stijn De Baerdemacker}
\affiliation{ Center for Molecular Modeling, Ghent University,Technologiepark 903, 9052 Zwijnaarde, Belgium}
\author{Sebastian Wouters}
\affiliation{ Center for Molecular Modeling, Ghent University,Technologiepark 903, 9052 Zwijnaarde, Belgium}
\author{Dimitri Van Neck}
\affiliation{ Center for Molecular Modeling, Ghent University,Technologiepark 903, 9052 Zwijnaarde, Belgium}
\begin{abstract}
The reduced density matrix is variationally optimized for the two-dimensional Hubbard model. Exploiting all symmetries present in the system, we have been able to study $6\times6$ lattices at various fillings and different values for the on-site repulsion, using the highly accurate but computationally expensive three-index conditions. To reduce the computational cost we study the performance of imposing the three-index constraints on local clusters of $2\times2$ and $3\times3$ sites. We subsequently derive new constraints which extend these cluster constraints to incorporate the open-system nature of a cluster on a larger lattice. The feasibility of implementing these new constraints is demonstrated by performing a proof-of-principle calculation on the $6\times6$ lattice. It is shown that a large portion of the three-index result can be recovered using these extended cluster constraints, at a fraction of the computational cost.
\end{abstract}
\pacs{03.67.Ac,03.65.Aa,71.10.Fd}
\keywords{variational,reduced density matrix,strongly correlated electrons}
\maketitle
\section{Introduction}
Although the underlying physical laws and mathematical formalism describing many-electron systems are fully understood, solving the associated quantum many-body problem remains a very challenging task. This is because the dimension of the associated Hilbert space increases exponentially with system size. To obtain results for correlated electron systems one therefore resorts to approximate techniques, which try to capture the relevant parts of the physics in the system. This paper deals with variational density matrix optimization (v2DM), a many-body method which removes the wave function from quantum mechanics by replacing it wth the two-body reduced density matrix (2DM), at the benefit of having nicer scaling properties. This appealing idea originates from the 1940's with Husimi \cite{husimi} and was brought to focus by Coleman in his excellent review paper \cite{coleman}. In subsequent years a lot of effort was put in the formulation of the problem \cite{garrod,kummer} and the first numerical calculations were performed \cite{fusco,garrod_mih_ros}. However, some disappointing results \cite{mihailovic,rosina} and the computational complexity of the problem caused activity in the field to drop. It wasn't until Nakata {\it et al.} \cite{nakata_first} and Mazziotti \cite{mazziotti} introduced a numerical technique called semidefinite programming to v2DM, that interest in the field was renewed. Since then a lot of progress has been made, both by developing better algorithms suited to the physics problem at hand \cite{primal_dual,maz_bp,maz_prl,nakata_last}, as well as by increasing the accuracy of the approximation by deriving new constraints \cite{zhao,mazz_T_con,dimi,qsep,shenvi}. A large variety of systems has been studied using v2DM, mostly atomic and molecular systems \cite{hammond,nakata_last,mazz_T_con,Gido_T_con,mazz_book,braams_book}, but also nuclear \cite{mihailovic,rosina}, and spin and lattice systems \cite{maz_hub,barthel,shenvi,nakata_last,hubbard_bv,gutz_sdp}. Recently it was observed that this standard approach fails to describe the dissociation limit of certain diatomic molecules \cite{helen_1}. This failure was shown to originate from the lack of size-extensivity of the method \cite{qsep,nakata_se}, and several ways to cure this behaviour have been set forth \cite{qsep,helen_2,nakata_li}.

This paper focuses on how the standard v2DM approach performs for the 2D Hubbard model, and proposes a computationally much cheaper way in which these results can be approximated. It should be noted that the proposed new conditions are approximations to the standard non-negativity constraints, and as such they do not intend to cure the non-size extensivity of the method.

The Hubbard model is a schematic model \cite{hubbard} developed to describe electron correlation in solids by means of the Hamiltonian:
\begin{equation}
\hat{H} = -t\sum_{\langle ij \rangle}\left(a^\dagger_{i\sigma}a_{j\sigma} + a^\dagger_{j\sigma}a_{i\sigma}\right) + U\sum_in^\uparrow_in^\downarrow_i~,\qquad\text{with}\qquad n^\sigma_i = a^\dagger_{i\sigma}a_{i\sigma}~,
\label{hub_ham}
\end{equation}
in which the sum $\langle ij \rangle$ goes over nearest neighbouring sites only, and using second-quantized notation where $a^\dagger_\alpha/a_\alpha$ creates/annihilates a fermion in a single-particle state $\alpha = \{i\sigma\}$ with $i$ the lattice label and $\sigma$ the spin projection ($\sigma = \pm\frac{1}{2}$), see {\it e.g.} \cite{bijbel,fetter_walecka}. The 2D version is of particular interest because it is thought to exhibit high-temperature superconductivity \cite{anderson,scalapino}. This simple model has a rich phase diagram and is therefore challenging to solve. This is due to the competition between the delocalizing first term (see Eq.~\ref{hub_ham}), which allows for hopping from one site to its neighbours, and the second term, which locally repels electrons that are on the same site. A lot of numerical studies have been carried out for this system, using various methods such as quantum Monte Carlo \cite{qmc_2dhub}, and the density matrix renormalization group (DMRG) \cite{xiang}. For an excellent overview of numerical studies performed on the 2D Hubbard model, we refer to the review by Scalapino \cite{scalapino}. In spite of all these numerical studies, surprisingly few reference results exist for the ground-state energy on finite size lattices. An earlier v2DM study focussed on the $4\times4$ lattice and was limited to the half-filled case \cite{nakata_2dhub}. In this paper we present results obtained by v2DM with the accurate three-index ($\mathcal{PQGT}'$) conditions (see Section \ref{intro}) on a $6\times6$ lattice at various fillings and for different values of the on-site repulsion $U$. In Section~\ref{intro} we give a short introduction to the general framework behind v2DM, and show how the symmetry present in the model can be exploited to obtain a substantial computational speed up. Subsequently we present and discuss the exact variational \emph{lower bound} results on the $6\times6$ lattice obtained using the computationally heavy $\mathcal{PQGT}'$ conditions and the much faster but less accurate two-index ($\mathcal{PQG}$) conditions. We compare these with an exact variational \emph{upper bound} obtained through a DMRG framework developed by one of us \cite{sebastian}. In Section~\ref{pure_cluster} we look for a way to recover three-index precision without losing computational efficiency, by imposing three-index constraints on local clusters. In Section~\ref{extcluster} we derive new constraints which extend these local cluster constraints in a way that reflects the open-system nature of a cluster. We demonstrate the feasibility of this approach by performing a proof-of-principle calculation imposing a subset of these constraints. It is shown that the computationally very expensive $\mathcal{PQGT}'$ results can be closely approximated by imposing only a limited subset of the extended clutser constraints, at a fraction of the computational cost.

\section{\label{intro}Theoretical Framework}
\subsection{\label{intro_2DM}Introduction to v2DM}
The Hamiltonian of a system interacting in up to two-body terms is:
\begin{equation}
\label{two_body_ham}
\hat{H}_{\alpha\beta;\gamma\delta} = \sum_{\alpha\beta}t_{\alpha\beta}a^\dagger_\alpha a_\beta + \frac{1}{4}\sum_{\alpha\beta\gamma\delta}V_{\alpha\beta;\gamma\delta}a^\dagger_\alpha a^\dagger_\beta a_\delta a_\gamma~,
\end{equation}
so the expectation value of the energy corresponding to an arbitrary $N$-particle state $\ket{\Psi^N}$ can be expressed using but the 2DM of that state:
\begin{equation}
\label{E_2DM}
E(\Gamma) = \mathrm{Tr}~\Gamma H^{(2)} = \frac{1}{4}\sum_{\alpha\beta\gamma\delta}\Gamma_{\alpha\beta;\gamma\delta}H^{(2)}_{\alpha\beta;\gamma\delta}~,
\end{equation}
with the 2DM defined as:
\begin{equation}
\Gamma_{\alpha\beta;\gamma\delta} = \bra{\Psi^N}a^\dagger_\alpha a^\dagger_\beta a_\delta a_\gamma \ket{\Psi^N}~,
\label{2DM}
\end{equation}
and the reduced two-body Hamiltonian:
\begin{equation}
\label{red_ham}
H^{(2)} = \frac{1}{N-1}\left[\delta_{\alpha\gamma}t_{\beta\delta}-\delta_{\alpha\delta}t_{\beta\gamma} -\delta_{\beta\gamma}t_{\alpha\delta}+\delta_{\beta\delta}t_{\alpha\gamma}\right] + V_{\alpha\beta;\gamma\delta}~.
\end{equation}
The key idea underpinning v2DM is to use Eq.~(\ref{E_2DM}) to determine the 2DM variationally. Once the approximate ground-state 2DM is found, we can extract all one- and two-body ground-state properties. This approach eliminates the need to reference the exponentially scaling wave function, and requires only the much more compact 2DM. A complication arises because the variation has to be performed over a limited set of 2DM's, {\it i.e.} those that are derivable from an ensemble of $N$-particle wave functions. This is known as the $N$-representability problem \cite{coleman}. Using the variational principle and the fact that the set of $N$-representable 2DM's is convex, one can derive a formal solution to the $N$-representability problem. This necessary and sufficient condition states that a $2$DM is $N$-representable if and only if 
\begin{equation}
\mathrm{Tr}~\Gamma~H^{(2)}_\nu \geq E^N_0\left(H^{(2)}_\nu\right)~,
\label{eq_dual_n_rep}
\end{equation}
for all $2$-particle Hamiltonians $H^{(2)}_\nu$, with corresponding ground-state energies $E^N_0(H^{(2)})$. These conditions are obviously not practically implementable, but can be used to derive a set of necessary constraints. Consider the class of positive Hamiltonians:
\begin{equation}
\hat{H} = B^\dagger B\qquad\text{with}\qquad B^\dagger = \sum_{\alpha\beta}g_{\alpha\beta}a^\dagger_\alpha a_\beta~.
\end{equation}
Inserting this class of Hamiltonians into Eq.~(\ref{eq_dual_n_rep}) leads to the inequality:
\begin{equation}
\sum_{\alpha\beta\gamma\delta}g_{\alpha\beta}\bra{\Psi^N}a^\dagger_\alpha a_\beta a^\dagger_\delta a_\gamma\ket{\Psi^N}g_{\gamma\delta} \geq 0~,
\end{equation}
from which the matrix-positivity condition \cite{garrod}
\begin{equation}
\mathcal{G}(\Gamma) \succeq 0 \qquad\text{with}\qquad\mathcal{G}(\Gamma)_{\alpha\beta;\gamma\delta}=\delta_{\beta\delta}\rho_{\alpha\gamma}-\Gamma_{\alpha\delta;\gamma\beta}~,
\end{equation}
follows. In this fashion one can derive two more necessary conditions \cite{coleman}, one is simply the non-negativity condition on the 2DM itself, another requires the non-negativity of:
\begin{equation}
\mathcal{Q}(\Gamma) \succeq 0 \qquad\text{with}\qquad \mathcal{Q}(\Gamma)_{\alpha\beta;\gamma\delta} = \bra{\Psi^N}a_\alpha a_\beta a^\dagger_\delta a^\dagger_\gamma\ket{\Psi^N}~.
\end{equation}
These three combined form the so-called two-index conditions, often referred to as $\mathcal{PQG}$.

Another class of constraints is derived using the positive Hamiltonian \cite{zhao,hammond,erdahl}:
\begin{equation}
\hat{H} = B^\dagger B + B B^\dagger~,
\end{equation}
in which the $B^\dagger$ is a three-body operator. Two conditions are derived in this way:
\begin{align}
\mathcal{T}_1\left(\Gamma\right) \succeq& 0 \qquad \text{using} \qquad B^\dagger = \sum_{\alpha\beta\gamma}t^1_{\alpha\beta\gamma}a^\dagger_\alpha a^\dagger_\beta a^\dagger_\gamma~,\qquad\text{and}\\
\label{T2p}\mathcal{T}_2'\left(\Gamma\right) \succeq& 0 \qquad \text{using} \qquad B^\dagger = \sum_{\alpha\beta\gamma}t^2_{\alpha\beta\gamma}a^\dagger_\alpha a^\dagger_\beta a_\gamma + \sum_\mu s_\mu a^\dagger_\mu~.
\end{align}
These are matrix-positivity constraints on respectively three-particle and two-particle-one-hole space, and are referred to as three-index constraints. Enforcing these conditions usually greatly improves the accuracy of the results, but at a considerable computational cost.

v2DM can now be formulated as the following constrained optimization problem:
\begin{equation}
\min_{\Gamma} \mathrm{Tr}~\Gamma H^{(2)}~,
\end{equation}
on the condition that
\begin{eqnarray}
\mathrm{Tr}~\Gamma &=& \frac{N(N-1)}{2}~,\\
\mathcal{L}(\Gamma) &\succeq& 0 \qquad \forall \mathcal{L} \in \{\mathcal{P,Q,G},\mathcal{T}_1,\mathcal{T}_2\}~.
\end{eqnarray}
Because the set of 2DM's over which the optimization is performed is too large, one obtains a variational lower bound to the energy, complementary to variational wave-function techniques where one finds an upper bound.

This problem is an instance of the class of semidefinite programs (SDP). This is a well studied optimization problem for which many algorithms have been developed. A number of these algorithms have been tailored to the specific case of v2DM, and they can be divided into two classes. On the one hand there are the interior-point methods \cite{nakata_first,primal_dual} which are very robust, but have a slow computational scaling of about $\mathcal{O}(M^{12})$, regardless of when two- and three-index conditions are imposed, where $M$ is the dimension of single-particle space. On the other hand there are first-order algorithms such as the boundary point method \cite{maz_bp,rendl}, which have a better scaling of $\mathcal{O}(M^9)$ when the three-index conditions are included, but these do not have the nice convergence properties of the interior point methods. In this study a boundary point method has been used.
\subsection{\label{symm}Symmetry adaptation of the constraints}
\begin{center}
\begin{table}
\begin{tabular}{|c|ccccc|ccc|ccc|}
\hline
$U$&\multicolumn{5}{c|}{$N=16$}&\multicolumn{3}{c|}{$N=14$}&\multicolumn{3}{c|}{$N=10$}\\
\hline
& exact &  $\mathcal{PQGT}'$& $\mathcal{PQGT}'$ \cite{nakata_2dhub}&$\mathcal{PQG}$& $\mathcal{PQG}$ \cite{nakata_2dhub}& exact& $\mathcal{PQGT}'$&$\mathcal{PQG}$&exact&$\mathcal{PQGT}'$&$\mathcal{PQG}$\\
\hline
4&-0.8514&-0.8607&-0.8607&-0.9343&-0.9341&-1.1246&-1.1486&-1.2664&-1.9581&-1.9595&-2.0106\\
8&-0.5329&-0.5476&-0.5476&-0.6604&-0.6603&-0.8478&-0.9124&-1.1162&-1.7510&-1.7597&-1.8806\\
12&-0.3745&-0.3909&*&-0.5054&*&-0.7180&-0.8024&-1.1050&-1.6455&-1.6626&-1.8284\\
16&-0.2882&-0.30101&*&-0.4073&*&-0.6475&-0.7409&-1.0136&-1.5843&-1.6085&-1.8014\\
\hline
\end{tabular}
\caption{\label{4x4}Ground-state energy per particle for the $4\times4$ lattice with different fillings. $\mathcal{PQG}$ and $\mathcal{PQGT}'$ results are compared with exact diagonalization result from \cite{fano,wardje}, and the previous v2DM reference results from \cite{nakata_2dhub}.} 
\end{table}
\end{center}
\begin{center}
\begin{table}
\begin{tabular}{|c|c|c|c|c|c|c|c|c|c|c|}
\hline
$U$& \multicolumn{2}{c|}{$N=36$} & \multicolumn{2}{c|}{$N=34$}& \multicolumn{2}{c|}{$N=30$} & $N=24$ & $N=18$ & $N=12$ &$ N=6$\\
\hline
&DMRG &$\mathcal{PQGT}'$&DMRG &$\mathcal{PQGT}'$&DMRG &$\mathcal{PQGT}'$&$\mathcal{PQGT}'$&$\mathcal{PQGT}'$&$\mathcal{PQGT}'$&$\mathcal{PQGT}'$\\
\hline
1	&  -1.3256     &-1.3399		&-1.4272       &	-1.4398		& -1.6710      &	-1.6775		&	-2.0981		&	-2.5533	   &  -2.9286	   &  -3.3131  \\
2	&	-1.1310     &-1.1543		&-1.2396       &	-1.2646		& -1.5036      &	-1.5194		&	-1.9720		&	-2.4606	   &  -2.8716		&	-3.2976  \\
3	&	-0.9708     &-0.9985		&-1.0835       &	-1.1212		& -1.3629      &	-1.3910	   &  -1.8694		&	-2.3851	   &  -2.8299		&	-3.2851  \\
4	&	-0.8409     &-0.8702		&-0.9577       &	-1.0066		& -1.2473      &	-1.2892		&	-1.7870		&	-2.3237	   &  -2.7945		&	-3.2751  \\
5	&	-0.7356     &-0.7654		&-0.8555       &	-0.9145		& -1.1540      &	-1.2094		&	-1.7212		&	-2.2738	   &  -2.7602		&	-3.2664  \\
6	&	-0.6501     &-0.6798		&-0.7746       &	-0.8395		& -1.0790      &	-1.1463		&	-1.6688		&	-2.2329	   &  -2.7369 		&	-3.2594  \\
7	&	-0.5803     &-0.6092		&-0.7080       &	-0.7776		& -1.0186      &	-1.0957		&	-1.6268		&	-2.1992		&	-2.7167		&	-3.2539  \\
8	&	-0.5226     &-0.5504		&-0.6538       &	-0.7261		& -0.9697      &	-1.0546		&	-1.5926		&	-2.1711		&	-2.7000		&	-3.2498  \\
9	&	-0.4744     &-0.5009		&-0.6083       &	-0.6828		& -0.9296      &	-1.0207		&	-1.5646		&	-2.1475		&	-2.6858		&	-3.2451  \\
10	&	-0.4339     &-0.4588		&-0.5707       &	-0.6459		& -0.8953      &	-0.9924		&	-1.5414		&	-2.1275		&	-2.6737		&	-3.2421  \\
11	&	-0.3994     &-0.4228		&-0.5388       &	-0.6144		& -0.8671      &	-0.9686		&	-1.5217		&	-2.1104		&	-2.6632		&	-3.2382  \\
12	&	-0.3696     &-0.3916		&-0.5114       &	-0.5871		& -0.8410      &	-0.9483		&	-1.5050		&	-2.0956		&	-2.6541		&	-3.2354  \\
13	&	-0.3439     &-0.3643		&-0.4872       &	-0.5633		& -0.8210      &	-0.9308		&	-1.4906		&	-2.0827		&	-2.6461		&	-3.2325  \\
14	&	-0.3214     &-0.3405		&-0.4671       &	-0.5424		& -0.8036      &	-0.9157		&	-1.4780		&	-2.0715		&	-2.6390		&	-3.2305  \\
15	&	-0.3016     &-0.3194		&-0.4489       &	-0.5239		& -0.7881      &	-0.9025		&	-1.4669		&	-2.0615		&	-2.6328		&	-3.2282  \\
\hline
\end{tabular}
\caption{\label{6x6}Ground-state energy per particle for $6\times6$ lattice with different fillings and on-site repulsion, obtained with $\mathcal{PQGT}'$ conditions, compared with DMRG results for $N=36$, 34 and 30.}
\end{table}
\end{center}

The 2D Hubbard model on a square lattice with periodic boundary conditions (PBC),  {\it i.e.} the 2D Hubbard model on a torus, has a lot of symmetry. This symmetry can be used to block diagonalize the constraint matrices $\mathcal{L}$, allowing for a reduction of the computational cost of the SDP. Full exploitation of this symmetry allows us to push our calculations to reasonably sized lattices, up to $20\times20$ when only two-index conditions are imposed, and up to $6\times6$ with the three-index conditions included. We will impose three different symmetries: spin symmetry, translational invariance and the point-group symmetry of the lattice.
\subsubsection{Spin Symmetry}
The invariance of the Hubbard Hamiltonian under rotations in spin space can be exploited by introducing the spin-averaged ensemble \cite{atomic}. The $\mathcal{G}$ condition, when defined in a spin-averaged way as:
\begin{equation}
\mathcal{G}(\Gamma)^S_{ab;cd} = \frac{1}{2\mathcal{S} + 1}\sum_{\mathcal{M}}\bra{\Psi^{\mathcal{SM}}}\left[a^\dagger_a \otimes \tilde{a}_b\right]^S \left(\left[a^\dagger_c \otimes \tilde{a}_d\right]^S\right)^\dagger\ket{\Psi^{\mathcal{SM}}}~,
\label{G_spin_averaged}
\end{equation}
breaks down into four blocks, one $S=0$ block and three degenerate $S = 1$ blocks. This does not change the scaling of the algorithm, but elementary matrix computations are performed 32 times faster. Analogous results are obtained for the other constraints.
\subsubsection{Translational invariance}
Because periodic boundary conditions are assumed, the Hamiltonian is invariant under lattice translations in both the $x$ and $y$ direction. It is straightforward to exploit this symmetry: by transforming the single-particle basis to quasi-momentum eigenstates:
\begin{equation}
\ket{k_xk_y;\sigma} = \frac{1}{L}\sum_{x,y=1}^Le^{ik_x x}e^{ik_y y}\ket{xy;\sigma}~,
\end{equation}
in which $L$ is the linear dimension of the lattice.
The constraint matrices then decompose into blocks with the same two-particle or particle-hole quasi-momentum $(K_x,K_y)$, {\it e.g.} the $S = 0$ part of the $\mathcal{G}$ matrix falls apart in $L^2$ blocks of dimension $L^2$. This \emph{does} change the scaling of the program. Instead of the $\mathcal{O}(L^{12})$ scaling of a matrix multiplication on the $\mathcal{G}$ matrix without translational invariance, we now have a scaling of $\mathcal{O}(L^8)$.
\subsubsection{Point group symmetry}
The final symmetry we exploit is the point-group symmetry ($C_{4v}$) of the lattice. Combined with translational invariance this forms the space group $P4mm$. There are three independent transformations which leave the Hamiltonian invariant:
\begin{align}
\ket{x,y} \rightarrow& \ket{-x , y}\qquad\text{reflection symmetry in the $x$-direction}\\
\ket{x,y} \rightarrow& \ket{x,-y}\qquad\text{reflection symmetry in the $y$-direction}\\
\ket{x,y} \rightarrow& \ket{y,x}\qquad\text{reflection symmetry along the diagonal}
\end{align}
These operations do not commute with the translation operator, which means they map different $(K_x,K_y)$ blocks in the constraint matrices onto each other. This can be used to reduce the number of blocks we have to store and perform calculations on. A problem arises in blocks that are mapped onto themselves by the symmetry operations. For consistency the symmetry then has to be enforced in these blocks by imposing linear constraints on the 2DM during the optimization.
\subsection{Reference results}
\begin{figure}
\centering
\includegraphics[scale=0.9]{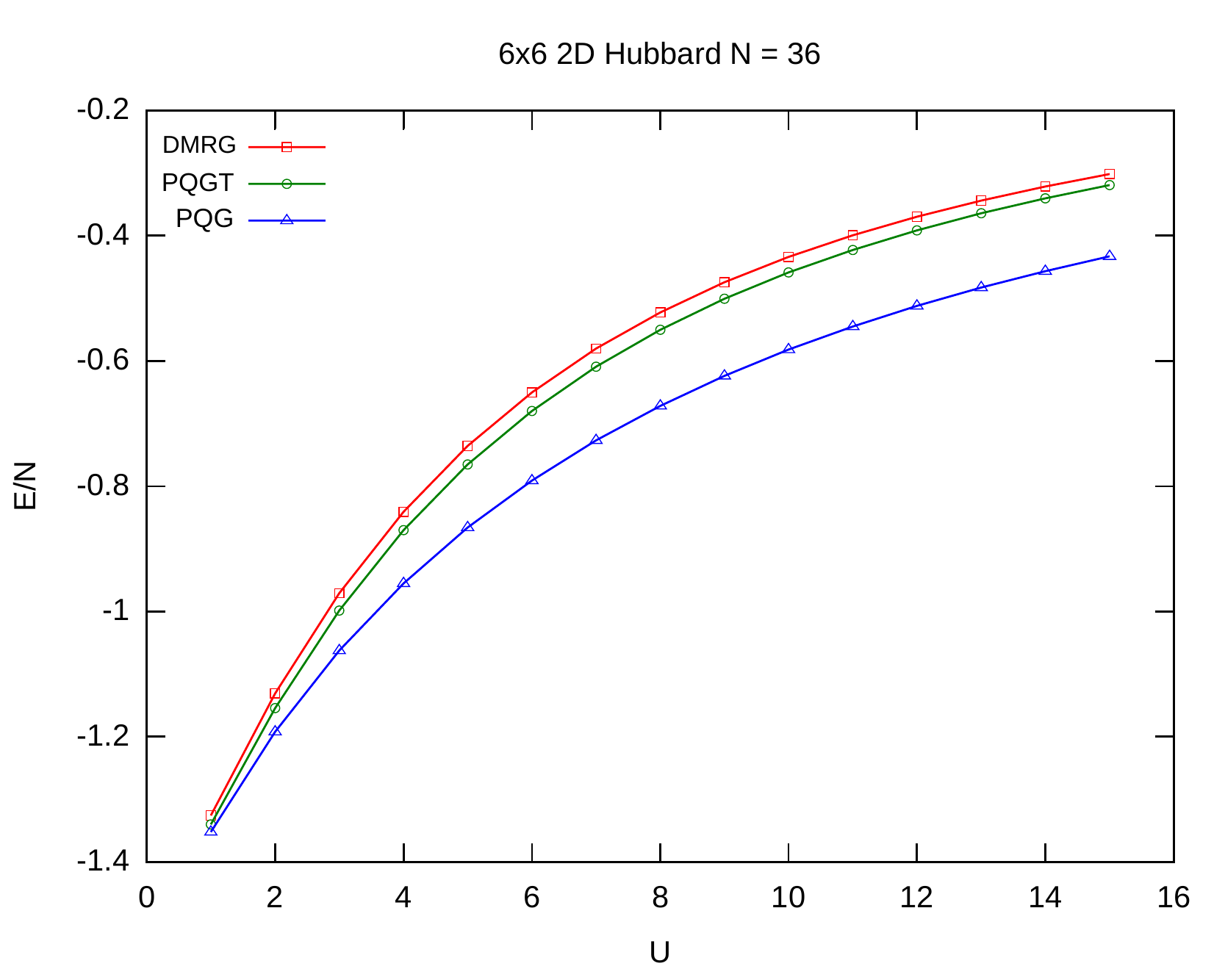}
\caption{\label{N36} Ground-state energy per particle as a function of on-site repulsion $U$ for 36 particles on a $6\times6$ lattice (cfr. Table~\ref{6x6}). DMRG results are compared with v2DM results using $\mathcal{PQGT}'$ and $\mathcal{PQG}$ conditions.}
\end{figure}
\begin{figure}
\centering
\includegraphics[scale=0.9]{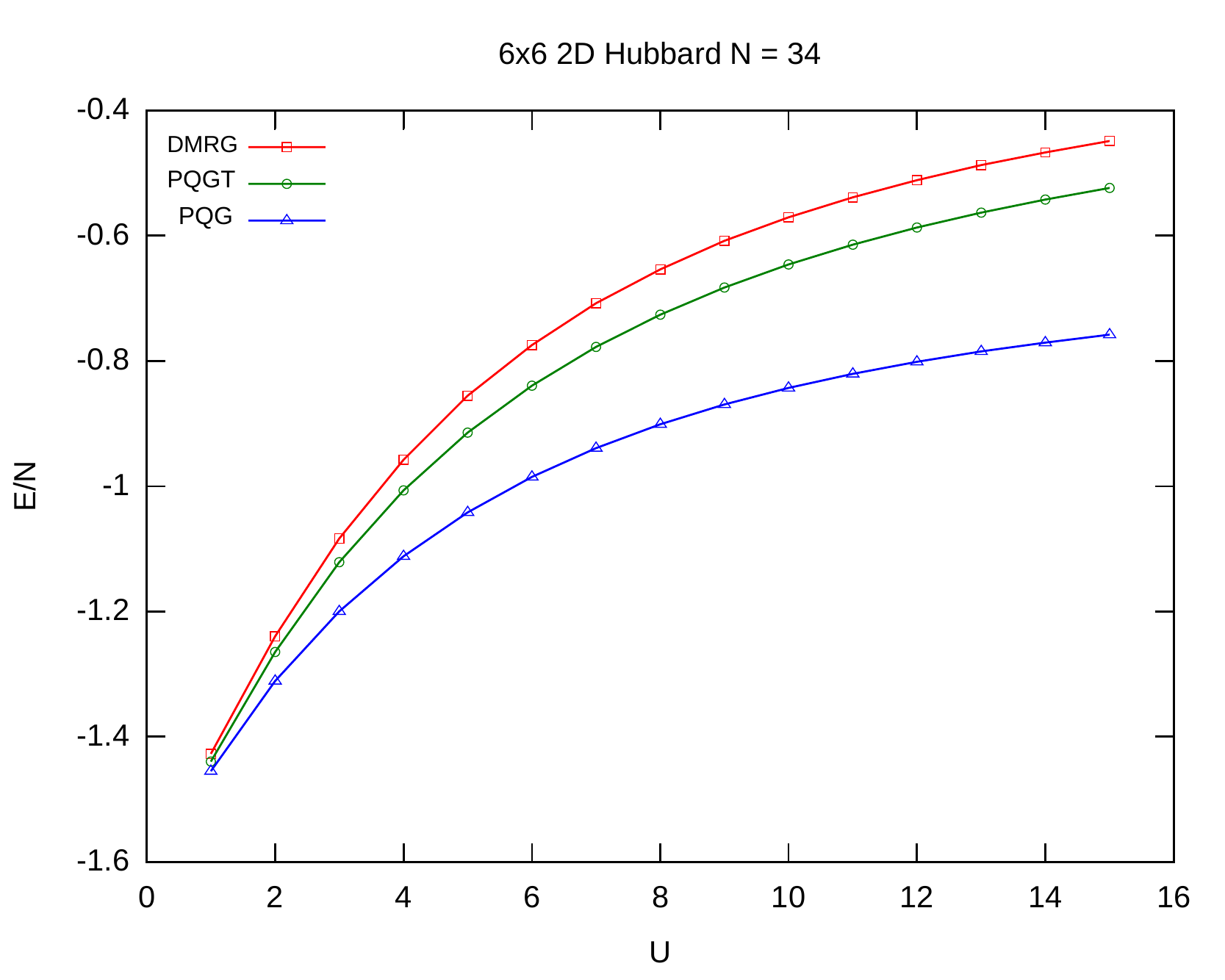}
\caption{\label{N34} Ground-state energy per particle as a function of on-site repulsion $U$ for 34 particles on a $6\times6$ lattice (cfr. Table~\ref{6x6}). DMRG results are compared with v2DM results using $\mathcal{PQGT}'$ and $\mathcal{PQG}$ conditions.}
\end{figure}
\begin{figure}
\centering
\includegraphics[scale=0.9]{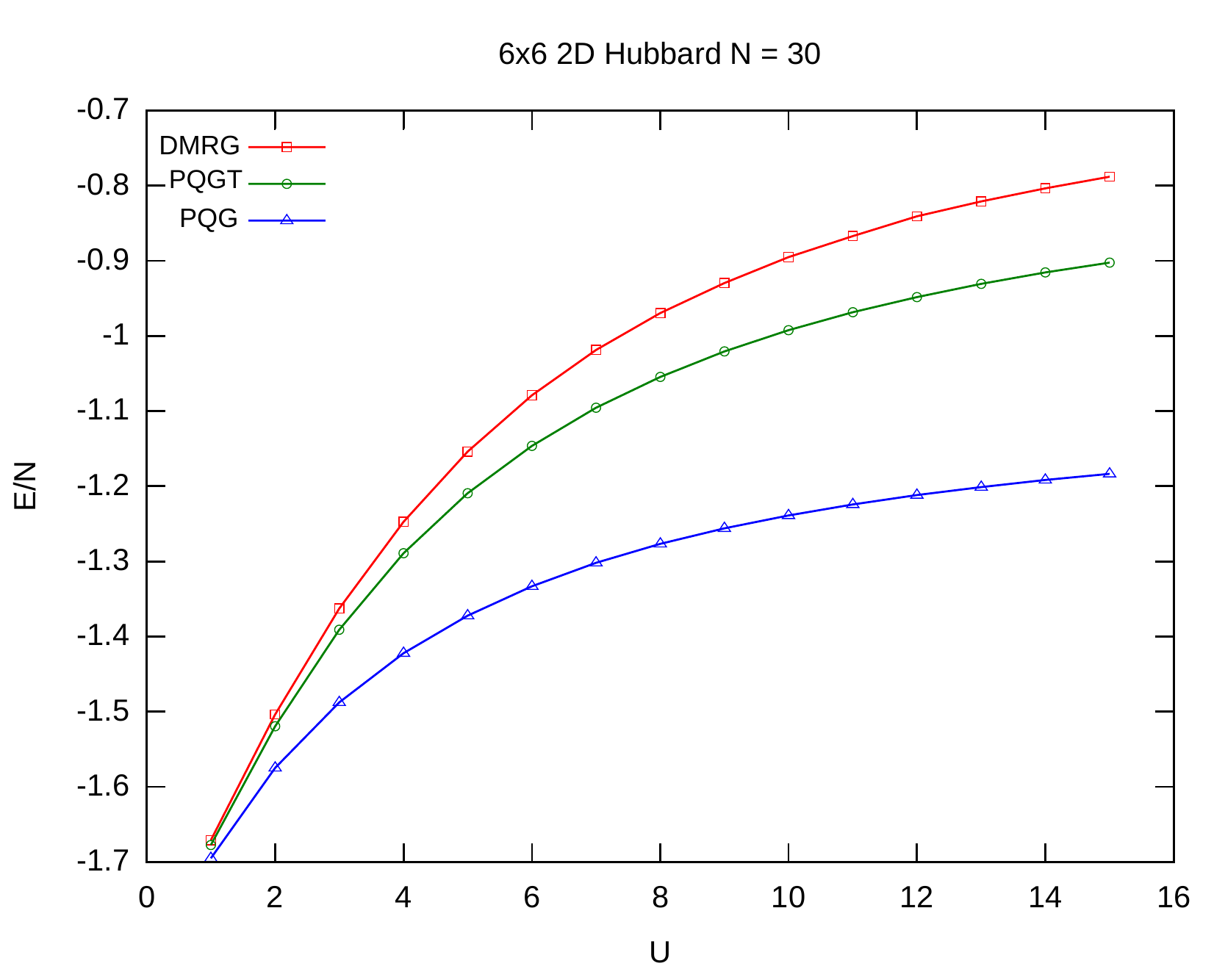}
\caption{\label{N30} Ground-state energy per particle as a function of on-site repulsion $U$ for 30 particles on a $6\times6$ lattice (cfr. Table~\ref{6x6}). DMRG results are compared with v2DM results using $\mathcal{PQGT}'$ and $\mathcal{PQG}$ conditions.}
\end{figure}

Using the method discussed in Sections \ref{intro_2DM} and \ref{symm} we have performed ground-state $\mathcal{PQGT}'$ calculations for the $6\times6$ Hubbard model, for different values of on-site repulsion $U$ and at various fillings. For reference, we compare our calculations of the energy per particle number on a $4\times4$ lattice with exact diagonalization results by Fano {\it et al.} \cite{fano} and an exact diagonalization program written by one of us \cite{wardje}. As can be seen in Table~\ref{4x4}, the $\mathcal{PQGT}'$ results are about $1\%$ below the exact result. As expected, just imposing the $\mathcal{PQG}$ constraints does not give satisfactory results. We also compare our v2DM results to those obtained by Anderson {\it et al.} in \cite{nakata_2dhub}, and it is shown that they correspond.

In Table~\ref{6x6} the $\mathcal{PQGT}'$ results for the ground-state energy per particle on a $6\times6$ lattice for different filling factors and on-site repulsion $U$ are given for future reference. They provide rigourous lower bounds on the energy. For the $6\times6$ lattice with 36, 34 and 30 particles we can compare with a variational upper bound on the energy obtained with a two-site $SU(2) \otimes U(1)$ spin- and particle number-adapted DMRG code developed by one of us \cite{sebastian}, with 7500 reduced renormalized basis states at each boundary. Note that this corresponds to a much larger number of effective renormalized basis states, as in a spin-adapted DMRG code only one basis state per multiplet is retained.

In Figures~\ref{N36}, \ref{N34}, and \ref{N30} the v2DM results obtained with both the $\mathcal{PQG}$ and $\mathcal{PQGT}'$ conditions are compared to the DMRG results. We notice that for half filling, the $\mathcal{PQGT}'$ and DMRG results are very close. $\mathcal{PQG}$ results, however, do not follow the same trend and deviate significantly from the other results. This discrepancy becomes even worse when we move away from half filling. There the $\mathcal{PQG}$ energies are almost twice the $\mathcal{PQGT}'$ results at large values of $U$. The gap between the upper bound provided by the DMRG results and the lower bound $\mathcal{PQGT}'$ results also becomes larger when we dope the lattice. This means that the $\mathcal{PQGT}'$ conditions perform worse when the system is more delocalized. The DMRG results also become less accurate when the system is delocalized, but are closer to the exact result. In Appendix \ref{error_est} an estimate of the error on the DMRG result is made by looking at the relation between the discarded weight and the energy. In the remainder of this paper we look for ways to bridge the gap between the $\mathcal{PQG}$ and $\mathcal{PQGT}'$ results while avoiding the considerable computational burden that is associated with the three-index constraints.

\section{\label{pure_cluster}Cluster constraints}
\begin{figure}
\centering
\includegraphics[scale=0.9]{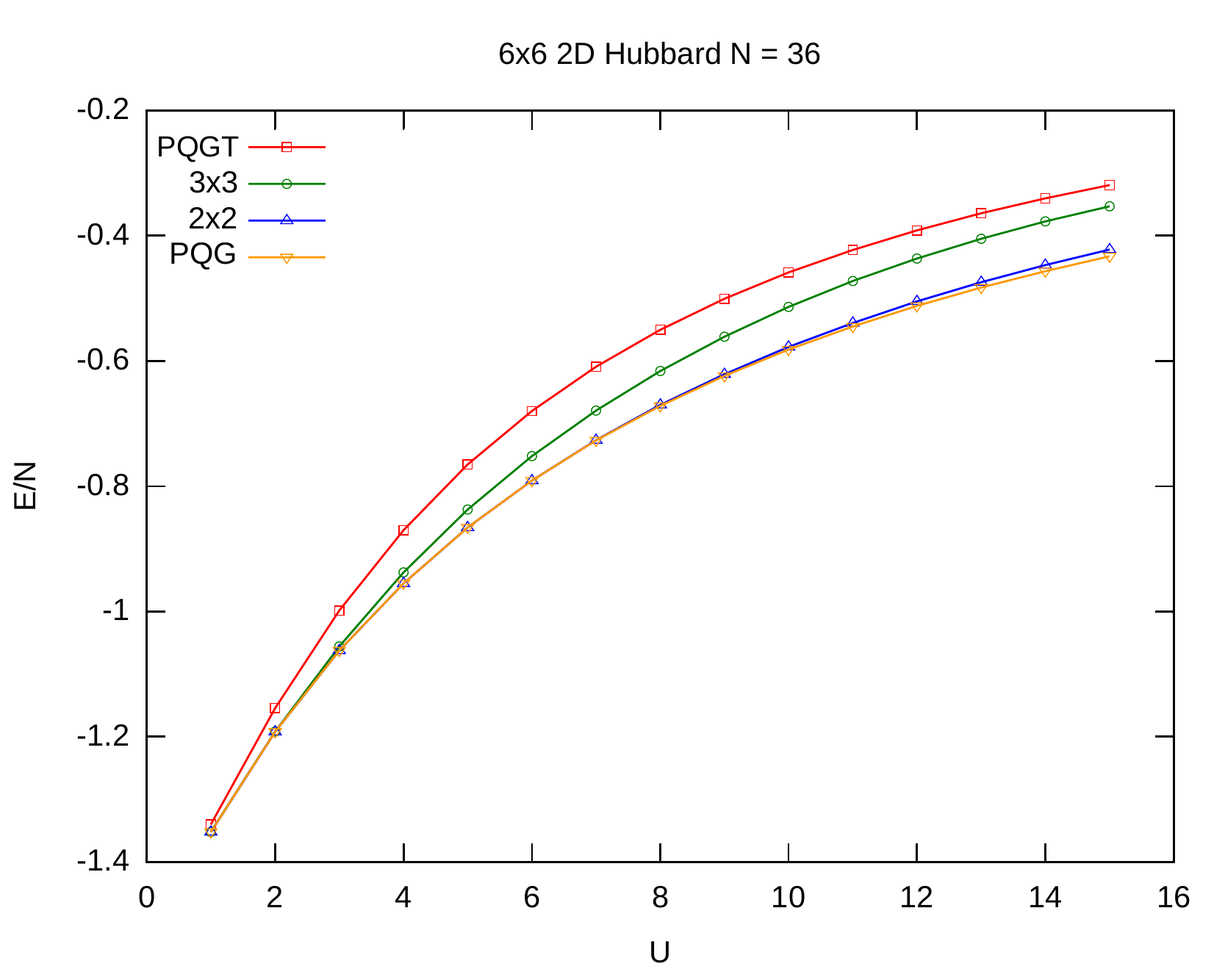}
\caption{\label{36c} Ground-state energy per particle as a function of on-site repulsion $U$ for 36 particles on a $6\times6$ lattice. Results of the three-index constraints imposed on clusters of $2\times2$ and $3\times3$ sites compared with the full $\mathcal{PQGT}'$ results and $\mathcal{PQG}$ results.}
\end{figure}

\begin{figure}
\centering
\includegraphics[scale=0.9]{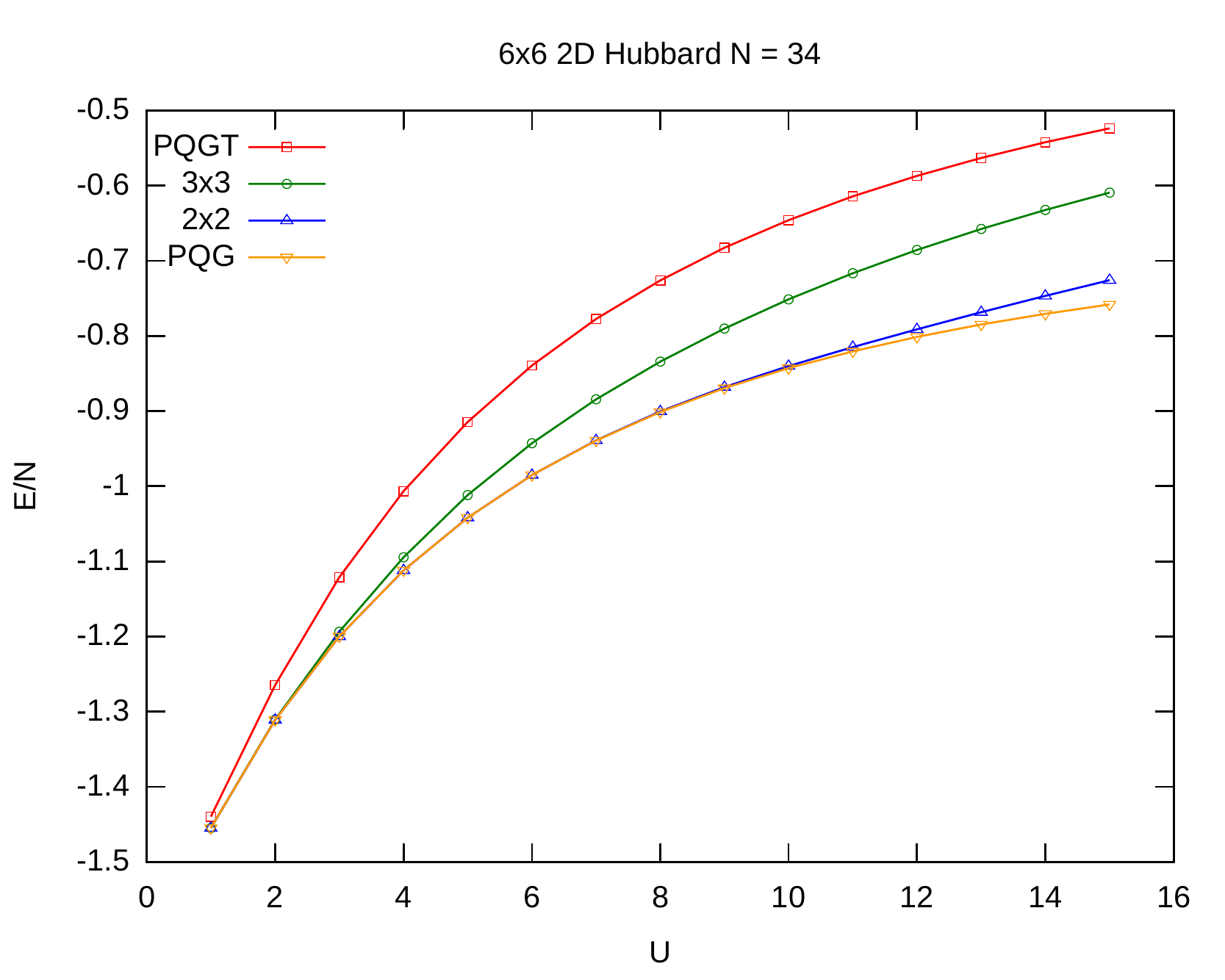}
\caption{\label{34c} Ground-state energy per particle as a function of on-site repulsion $U$ for 34 particles on a $6\times6$ lattice. Results of the three-index constraints imposed on clusters of $2\times2$ and $3\times3$ sites compared with the full $\mathcal{PQGT}'$ results and $\mathcal{PQG}$ results.}
\end{figure}

\begin{figure}
\centering
\includegraphics[scale=0.9]{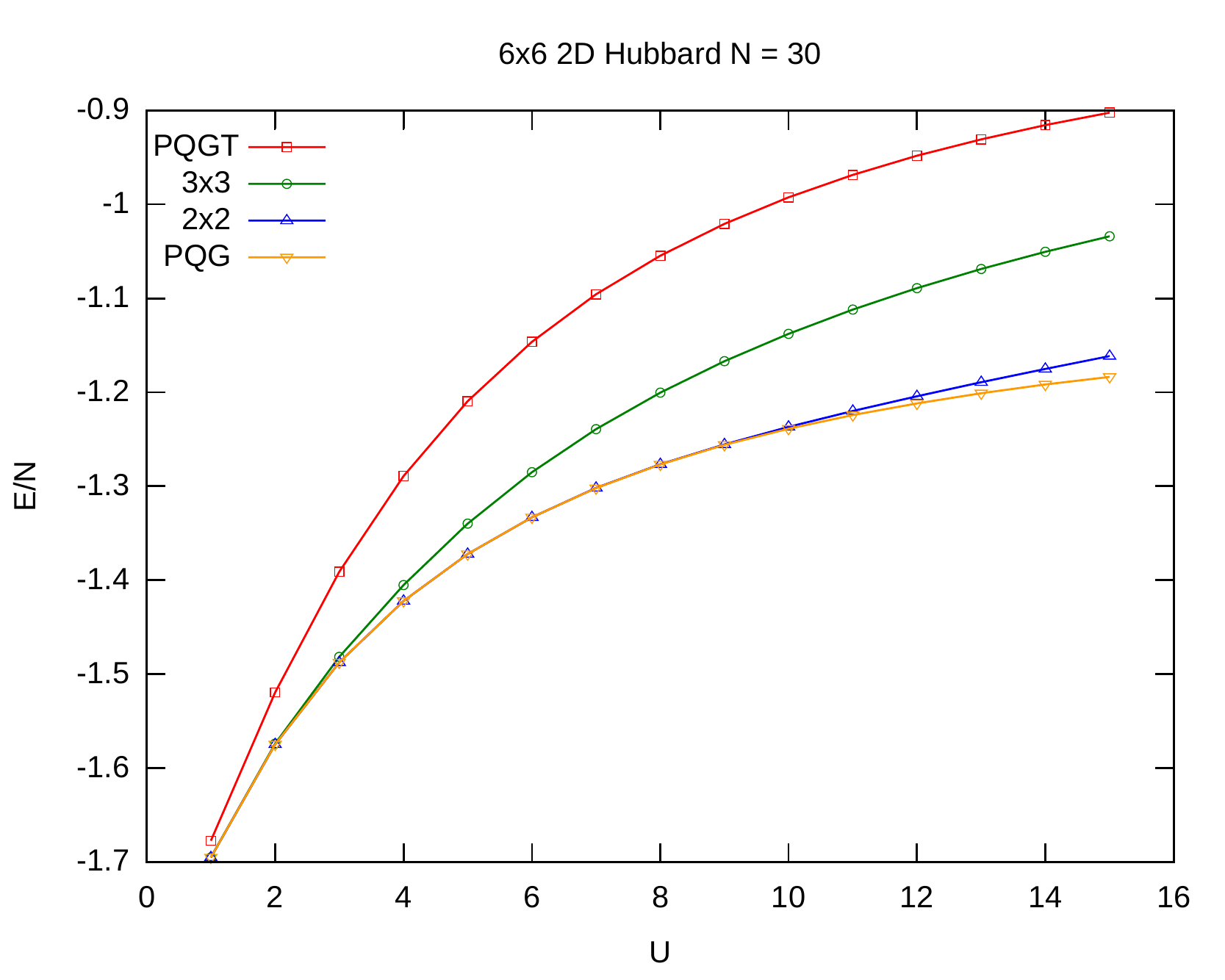}
\caption{\label{30c} Ground-state energy per particle as a function of on-site repulsion $U$ for 30 particles on a $6\times6$ lattice. Results of the three-index constraints imposed on clusters of $2\times2$ and $3\times3$ sites compared with the full $\mathcal{PQGT}'$ results and $\mathcal{PQG}$ results.}
\end{figure}

As the 2D Hubbard Hamiltonian only connects nearest neighbouring sites, it is reasonable to assume that the local correlations, occuring between adjacent sites, are particularly important. Various many-body methods take advantage of this observation and treat the local degrees of freedom on a higher level than the rest of the system \cite{dmft,review_cluster,dmet}. In a related approach we propose to impose the three-index conditions on smaller local clusters of $2\times 2$ or $3\times 3$ sites, while the full system is treated on a $\mathcal{PQG}$ level. This means we impose non-negativity only on those blocks of the $\mathcal{T}_1$ and $\mathcal{T}_2'$ matrices for which all single-particle indices are on the local cluster. A problem with imposing these local constraints is that we have to Fourier transform the 2DM back from the quasi-momentum basis to the site basis, which leads to some overhead. However, the matrix computations are the bottleneck of the program, and these become cheaper, since only subcluster matrices are considered. As a result the algorithm runs much faster than the full $6\times6$ $\mathcal{PQGT}'$ calculations. Because translational invariance is still imposed, only one cluster constraint has to be taken into account. In Figures~\ref{36c}, \ref{34c} and \ref{30c} it is shown what the improvement upon $\mathcal{PQG}$ is when implementing the three-index constraints on clusters of $2\times2$ and $3\times3$ sites. The $2\times2$ clusters seem to be too small, and the constraints only become active for larger values of on-site repulsion $U$. At half filling the three-index constraints on $3\times3$ clusters do a decent job, they bridge about half of the gap between $\mathcal{PQG}$ and $\mathcal{PQGT}'$. For 34 particles the performance is similar, whereas for 30 particles the result gets slightly worse, but the cluster constraints still manage to recover a substantial part of the correlation. This decrease in accuracy when moving away from half filling is not surprising, because one would not expect local constraints to be as effective in more delocalized systems.

In fact it is a bit naive to impose the three-index constraints on clusters, because one implicitly assumes that the subsystem is a closed system, {\it i.e.} that particle number is conserved on the cluster. The full Hamiltonian (\ref{hub_ham}), however, allows for the hopping of particles between the cluster and the rest of the system. We have access to the full system 2DM, which contains all the information about the communication between the cluster and the rest of the system. This observation suggests that there must be a way to extend the three-index cluster constraints to include the open system characteristics of the cluster, using the full system 2DM. In the next Section we derive new constraints which do exactly that, and we demonstrate the feasibility of the approach by implementing them in a proof-of-principle calculation.

\section{\label{extcluster}Extended cluster constraints}
\subsection{\label{eT1}The extended $\mathcal{T}_1$ constraint: $e\mathcal{T}_1$}
To simplify the presentation the constraints are derived in a spin-uncoupled form. The actual implementation of the constraints, however, was performed in a spin-coupled fashion. This changes nothing to the generality of the derived conditions or to the quality of the obtained results. As explained in Section~\ref{intro} the $\mathcal{T}_1$ condition is derived by demanding the non-negativity of the class of Hamiltonians:
\begin{equation}
\hat{H} = \left\{B^\dagger,B\right\}~,\qquad\text{where}\qquad B^\dagger = \sum_{\alpha\beta\gamma}t^1_{\alpha\beta\gamma}a^\dagger_\alpha a^\dagger_\beta a^\dagger_\gamma~.
\end{equation}
When we limit the $\mathcal{T}_1$ condition to a cluster we impose the non-negativity of a subclass of Hamiltonians generated by the three-particle operator:
\begin{equation}
B^\dagger = \sum_{abc}t^1_{abc}a^\dagger_a a^\dagger_b a^\dagger_c~,
\end{equation}
in which Latin indices $a$ are used to denote cluster states, overlined Latin indices $\bar{a}$ to denote the rest of the system and Greek indices $\alpha$ for general states.
A better approximation to the full system $B^\dagger$ is obtained if one allows one or two creation operators to be outside the cluster:
\begin{equation}
B^\dagger = \sum_{abc}t^1_{abc}a^\dagger_a a^\dagger_b a^\dagger_c+\sum_{ab\bar{c}}t^1_{ab\bar{c}}a^\dagger_a a^\dagger_b a^\dagger_{\bar{c}}+\sum_{\bar{a}\bar{b}c}t^1_{\bar{a}\bar{b}c}a^\dagger_{\bar{a}} a^\dagger_{\bar{b}} a^\dagger_c~.
\label{ecT1_full}
\end{equation}
In this way it is clear that the hopping in and out of the cluster is included. However, we want to construct a condition which only depends on the cluster indices. This can be accomplished by factorizing the terms in Eq.~(\ref{ecT1_full}) which contain both cluster and non-cluster indices:
\begin{align}
t^1_{ab\bar{c}} =& x_{ab} p_{\bar{c}}~,\\
t^1_{\bar{a}\bar{b}c} =& g_{\bar{a}\bar{b}} z_{c}~.
\end{align}
The new $B^\dagger$ operator only depends on cluster indices, a one-particle ($p^\dagger = \sum_{\bar{c}} p_{\bar{c}}a^\dagger_{\bar{c}}$) and a two-particle state ($g^\dagger = \sum_{\bar{a}\bar{b}}g_{\bar{a}\bar{b}}a^\dagger_{\bar{a}}a^\dagger_{\bar{b}}$) outside of the cluster:
\begin{equation}
B^\dagger = \sum_{abc}t^1_{abc}a^\dagger_a a^\dagger_b a^\dagger_c+\sum_{ab}x_{ab}a^\dagger_a a^\dagger_b p^\dagger+\sum_{c}z_c g^\dagger a^\dagger_c~.
\label{ecT1_red}
\end{equation}
With this $B^\dagger$, a new cluster constraint can be constructed, containing the three-index $\mathcal{T}_1$ and extending it with two- and one-index terms:
\begin{equation}
\left(
\begin{matrix}
(\mathcal{T}_1)_{abc;dez} & X(p)_{abc;st} & Z(g)_{abc;v}\\
X^\dagger(p)_{mn;dez} & X^{(2)}(p)_{mn;st}&(XZ)(p,g)_{mn;v}\\
Z^\dagger(g)_{u;dez} & (XZ)^\dagger(p,g)_{u;st}&Z^{(2)}(g)_{u;v}
\end{matrix}
\right)
\succeq 0~, \end{equation}
where the different tensors are defined as:
\begin{align}
X(p)_{abc;st} =& \langle a^\dagger_a a^\dagger_b a^\dagger_c p a_t a_s + p a_t a_s a^\dagger_a a^\dagger_b a^\dagger_c\rangle~,\\
Z(g)_{abc;v} =& \langle a^\dagger_a a^\dagger_b a^\dagger_c a_v  g + a_v g a^\dagger_a a^\dagger_b a^\dagger_c\rangle~,\\
X^{(2)}(p)_{mn;st} =&  \langle a^\dagger_m a^\dagger_n p^\dagger p a_t a_s + p a_t a_s a^\dagger_n a^\dagger_m p^\dagger\rangle~,\\
XZ(p,g)_{mn;v} =&  \langle a^\dagger_m a^\dagger_n p^\dagger a_v g + a_v g a^\dagger_n a^\dagger_m p^\dagger\rangle~,\\
Z^{(2)}(g)_{u;v} =&  \langle g^\dagger a^\dagger_u a_v g + a_v g g^\dagger a^\dagger_u\rangle~.
\end{align}
All these extra terms can be constructed using the full system 2DM and the knowledge of some predefined one- and two-particle states $p$ and $g$ outside the cluster.
\subsection{\label{eT2}The extended $\mathcal{T}_2'$ constraint: $e\mathcal{T}_2'$}
In an analogous way as for the $\mathcal{T}_1$ we can construct extensions which include cluster-system information to the $\mathcal{T}_2'$ condition. The two-particle-one-hole operator which constructs the full system $\mathcal{T}_2'$ (see Eq.~(\ref{T2p})) can be approximated on the cluster by:
\begin{equation}
B^\dagger = \sum_{abc}t^2_{abc}a^\dagger_a a^\dagger_b a_c + \sum_m s_m a^\dagger_m + \sum_{ab}x_{ab}a^\dagger_a a^\dagger_b h + \sum_{bc} y_{bc}p^\dagger a^\dagger_b a_c + \sum_{c}z_c g^\dagger a_c + \sum_{a}r_{a}a^\dagger_a d^\dagger ~,
\label{ecT2_red}
\end{equation}
in which the different states outside of the cluster are defined as:
\begin{align}
h = \sum_{\bar{c}}h_{\bar{c}}a_{\bar{c}}~,\qquad&\qquad p^\dagger = \sum_{\bar{a}}p_{\bar{a}}a^\dagger_{\bar{a}}~,\\
g^\dagger = \sum_{\bar{a}\bar{b}}g_{\bar{a}\bar{b}}a^\dagger_{\bar{a}}a^\dagger_{\bar{b}}~,\qquad&\qquad d^\dagger = \sum_{\bar{b}\bar{c}}d_{\bar{b}\bar{c}}a^\dagger_{\bar{b}}a_{\bar{c}}~.
\end{align}
As before this leads to a new non-negativity constraint which includes the old $\mathcal{T}_2'$ on the cluster and extends it with four new terms.
{\small
\begin{equation}
\left(
\begin{matrix}
(\mathcal{T}_2)_{abc;dez} &  P_{abc;w} & X(h)_{abc;st} & Y(p)_{abc;pq} & Z(g)_{abc;v} & R(d)_{abc;l}\\
P^\dagger_{r;dez} & P^{(2)}_{r;w} &PX(h)_{r;st}& PY(p)_{r;pq}& PZ^\dagger(g)_{r;v}& PR^\dagger(d)_{r;l} \\ 
X^\dagger(h)_{mn;dez} & PX^\dagger(h)_{mn;w}& X^{(2)}(h)_{mn;st} & XY(h,p)_{mn;pq} & XZ(h,g)_{mn;v} & XR(h,d)_{mn;l} \\
Y^\dagger(p)_{xy;dez} &  PY^\dagger(p)_{xy;w}&XY^\dagger(h,p)_{xy;st} & Y^{(2)}(p)_{xy;pq} & YZ(p,g)_{xy;v}& YR(p,d)_{xy;l}\\
Z^\dagger(g)_{u;dez} & PZ^\dagger(g)_{u;w}  & XZ^\dagger(h,g)_{u;st}& YZ^\dagger(p,g)_{u;pq}& Z^{(2)}(g)_{u;v} & ZR(g,d)_{u;l}\\
R^\dagger(d)_{k;dez} &  PR^\dagger(d)_{k;w}& XR^\dagger(h,d)_{k;st}& YR^\dagger(p,d)_{k;pq}& ZR^\dagger(g,d)_{k;v}& R^{(2)}(d)_{k;l}
\end{matrix}
\right)
\succeq 0 
\end{equation}
}
in which the extra tensor terms are defined as:
\begin{align*}
X(h)_{abc;st} = \langle a^\dagger_a a^\dagger_b a_c h^\dagger a^\dagger_t a_s + h^\dagger a^\dagger_t a_s a^\dagger_a a^\dagger_b a_c \rangle~,&\quad Y(p)_{abc;pq} = \langle a^\dagger_a a^\dagger_b a_c a^\dagger_q a_p p + a^\dagger_q a_p p a^\dagger_a a^\dagger_b a_c  \rangle~,\\
Z(g)_{abc;v} = \langle a^\dagger_a a^\dagger_b a_c a^\dagger_v g + a^\dagger_v g a^\dagger_a a^\dagger_b a_c  \rangle~,&\quad R(d)_{abc;l} = \langle a^\dagger_a a^\dagger_b a_c d a_l + d a_l a^\dagger_a a^\dagger_b a_c  \rangle~,\\
X^{(2)}(h)_{mn;st} =\langle a^\dagger_m a^\dagger_n h h^\dagger a_t a_s  + h^\dagger a_t a_s a^\dagger_m a^\dagger_n h  \rangle~,&\quad XY(h,p)_{mn;pq} =\langle a^\dagger_m a^\dagger_n h a^\dagger_q a_p p + a^\dagger_q a_p p a^\dagger_m a^\dagger_n h \rangle~,\\
XZ(h,g)_{mn;v} =\langle a^\dagger_m a^\dagger_n h a^\dagger_v g + a^\dagger_v g a^\dagger_m a^\dagger_n h \rangle~,&\quad XR(h,d)_{mn;l} =\langle a^\dagger_m a^\dagger_n h  d a_l + d a_l a^\dagger_m a^\dagger_n h \rangle~,\\
XP(h)_{mn;w} =\langle a^\dagger_m a^\dagger_n h  a_w \rangle~,&\quad Y^{(2)}(p)_{xy;pq} =\langle p^\dagger a^\dagger_x a_y a^\dagger_q a_p p +   a^\dagger_q a_p p p^\dagger a^\dagger_x a_y \rangle\\
YZ(p,g)_{xy;v} =\langle p^\dagger a^\dagger_x a_y a^\dagger_v g +   a^\dagger_v g p^\dagger a^\dagger_x a_y \rangle~,&\quad YR(p,d)_{xy;v} =\langle p^\dagger a^\dagger_x a_y d a_l +   d a_l p^\dagger a^\dagger_x a_y \rangle\\
YP(p)_{xy;w} =\langle p^\dagger a^\dagger_x a_y a_w\rangle~,&\quad Z^{(2)}(g)_{u;v} = \langle g^\dagger a_u a_v g + a_v g g^\dagger a_u \rangle\\
ZR(g,d)_{u;l} = \langle g^\dagger a_u d a_l + d a_l g^\dagger a_u \rangle~,&\quad ZP(g)_{u;w} = \langle g^\dagger a_u a_w \rangle~,\\
R^{(2)}(d)_{k;l} = \langle a^\dagger_k d^\dagger d a_l  + d a_l a^\dagger_k d^\dagger \rangle~,&\quad RP(d)_{k;w} = \langle a^\dagger_k d^\dagger a_w \rangle~.
\end{align*}
which can all be constructed using the full system 2DM and the different states ($h,p,g$ and $d$) outside of the cluster.
\subsection{\label{optim_con}Finding the optimal constraint}

\begin{figure}
\centering
\includegraphics[scale=0.9]{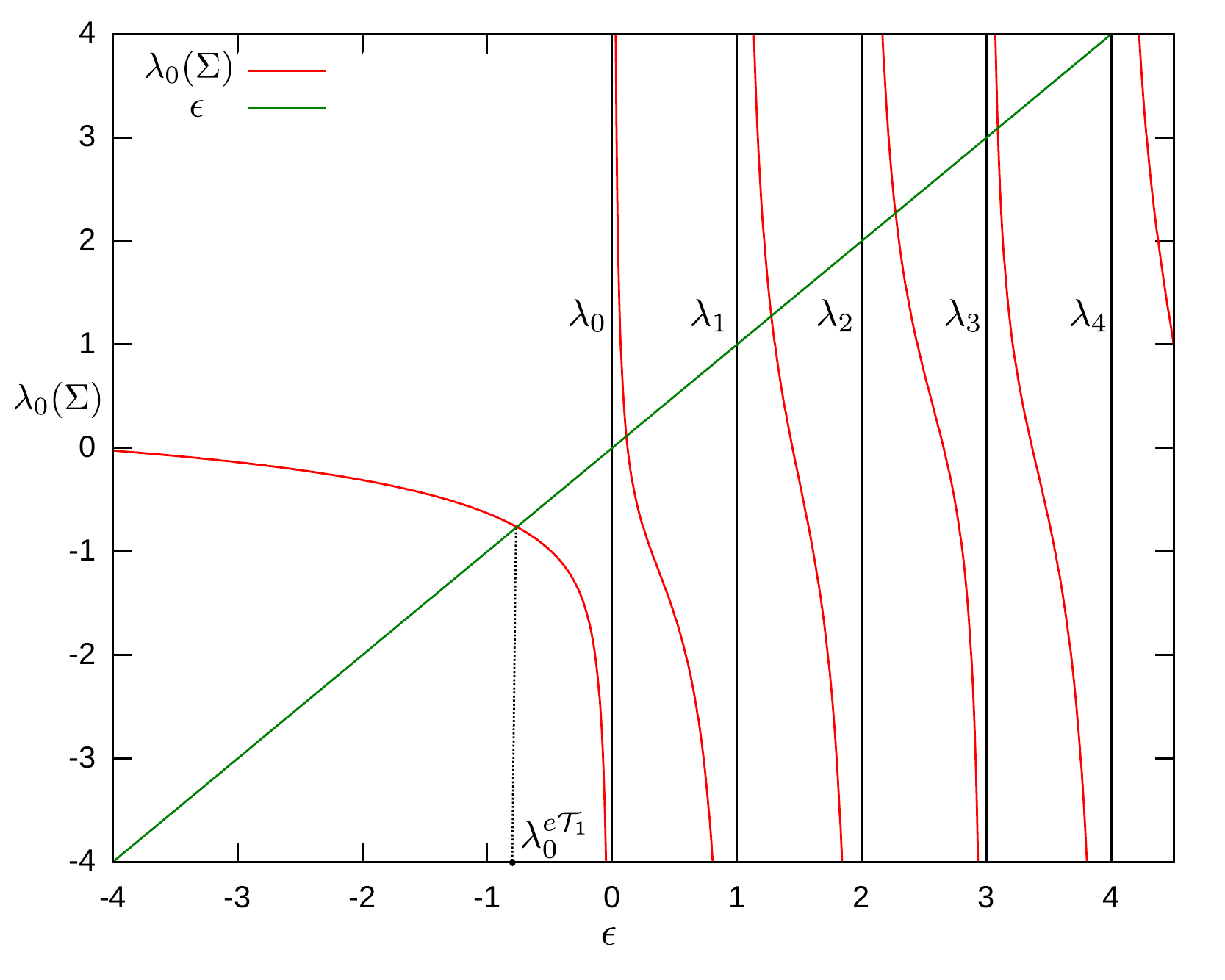}
\caption{\label{bisection} Illustration of the self-consistent eigenvalue problem (\ref{sc_eig}). The solution is found when the lowest eigenvalue of $\Sigma(\epsilon)$ ($\lambda_0(\Sigma)$) equals $\epsilon$. The black vertical lines are poles in $\Sigma(\epsilon)$ and correspond to the precalculated eigenvalues of the $\mathcal{T}_1$ matrix on the cluster.}
\end{figure}

In Sections~\ref{eT1} and \ref{eT2} we derived new constraints which include and extend the three-index constraints imposed on local clusters on the lattice. These new constraints depend on states defined outside of the cluster, giving rise to additional complications. In this Section we explore a method to choose these states, optimizing the class of positive Hamiltonians generated by Eqs.~(\ref{ecT1_red}) and (\ref{ecT2_red}) to make the constraints as strict as possible. In what follows we limit ourselves to the extended constraints with only one index outside of the cluster, {\it i.e.} for the $e\mathcal{T}_1$ there is only the dependence on the single-particle state $p$, whereas the $e\mathcal{T}_2'$ depends on the single-hole state $h$ and the single-particle state $p$.

The problem we discuss is the following: given a 2DM $\Gamma$ from some previous calculation, for what state $p$, outside of the cluster, is the $e\mathcal{T}_1(\Gamma)$ constraint  maximally violated. We have chosen to define the most violated constraint as the $e\mathcal{T}_1$ which has the lowest eigenvalue, {\it i.e.} we optimize the following cost function:
\begin{equation}
\phi^{e\mathcal{T}_1}(p) = \lambda^{e\mathcal{T}_1}_0(p) + \mu (|p|^2 - C)^2~,
\end{equation}
as a function of the single-particle state $p$, in which $\mu$ scales the quadratic potential that makes sure the $p$ vector is normalized, and $\lambda_0^{e\mathcal{T}_1}(p)$ is the lowest eigenvalue of the $e\mathcal{T}_1$ matrix for the current value of the state $p$. The gradient of this cost function can be evaluated analytically using the Hellman-Feynman theorem \cite{hellman_feynman}:
\begin{equation}
\frac{\partial \lambda_0^{e\mathcal{T}_1}}{\partial p_{\bar{z}}} = \bra{\Psi_0}\left(\frac{\partial e\mathcal{T}_1(p)}{\partial p_{\bar{z}}}\right)\ket{\Psi_0}~,\qquad\text{with}\qquad e\mathcal{T}_1(\Gamma)\ket{\Psi_0} = \lambda_0^{e\mathcal{T}_1}\ket{\Psi_0}~.
\end{equation}
The cost function can then be optimized using a simple non-linear conjugate gradient algorithm \cite{jrs}. For the construction of an efficient conjugate gradient algorithm it is essential to have a fast evaluation of the lowest eigenvalue and eigenvector of the constraint with a certain state $p$. This can be achieved because the largest block, the $\mathcal{T}_1$ on the cluster, remains unchanged during the optimization. If we prediagonalize the large block on the cluster, we need a fast method to solve the following problem:
\begin{equation}
\left[
\begin{matrix}
\lambda_1 &0&\ldots& 0 & b_{11} & \ldots & b_{1m}\\
0 &\lambda_2& \ldots& 0 & b_{21} & \ldots & b_{2m}\\
0 & 0 &\ddots  &0 & \vdots & \vdots & \vdots\\
0 & 0 & \ldots &\lambda_n & b_{n1}&\ldots&b_{nm}\\
b_{11}& b_{21} & \ldots &b_{n1} & d_{11} &\ldots & d_{1m}\\
\vdots& \vdots & \ldots &\vdots & \vdots &\ldots & \vdots\\
b_{1m} & b_{2m} & \ldots & b_{nm} & d_{m1} & \ldots & d_{mm}
\end{matrix}
\right]
\left[
\begin{matrix}
x_1\\
x_2\\
\vdots\\
x_n\\
y_1\\
\vdots\\
y_m
\end{matrix}
\right]
=
\epsilon
\left[
\begin{matrix}
x_1\\
x_2\\
\vdots\\
x_n\\
y_1\\
\vdots\\
y_m
\end{matrix}
\right]~,
\end{equation}
in which the $b$ and $d$ coefficients represent the extensions to the cluster constraint, which depend on the variable $p$, the $\lambda_i$'s are the precalculated eigenvalues of the $\mathcal{T}_1$ on the cluster, and $\epsilon$ is the lowest eigenvalue of the total matrix.
This leads to $n+m$ linear equations:
\begin{align}
\lambda_i x_i + \sum_{j=1}^m b_{ij} y_j =& \epsilon x_i\qquad\text{for}\qquad i = 1\ldots n~,\\
\sum_{i = 1}^n b_{ij} x_i + \sum_j d_{ij} y_i =& \epsilon y_j\qquad\text{for}\qquad j = 1\ldots m~.
\end{align}
From the first $n$ equations the $x_i$'s can be eliminated and expressed as a function of the $y$'s:
\begin{equation}
x_i = \frac{1}{\epsilon - \lambda_i}\sum_{k=1}^m {b_{ik}}y_k~.
\end{equation}
When this is substituted into the remaining $m$ equations we get the following self-consistent eigenvalue problem:
\begin{equation}
\sum_{k=1}^m\Sigma(\epsilon)_{jk}~y_k = \epsilon y_j~,\quad\text{for}\quad j = 1\ldots m~,\quad\text{with}\quad \Sigma(\epsilon)_{jk} = \sum_{i=1}^n\frac{b_{ij}b_{ik}}{\epsilon - \lambda_i} + d_{kj}~.
\label{sc_eig}
\end{equation}
These equations can be solved easily and quickly using the bisection method. As shown in Figure~\ref{bisection}, the lowest eigenvalue of the full matrix $\lambda_0^{e\mathcal{T}_1}$ is always below the lowest eigenvalue of the block matrix (the black vertical lines). This means one can easily bracket the lowest eigenvalue $\lambda_0^{e\mathcal{T}_1}$ and improve the approximation to it using the bisection method at the cost of just a couple of diagonalizations of an $m\times m$ matrix.
\subsection{Results}
\begin{figure}
\centering
\includegraphics[scale=0.9]{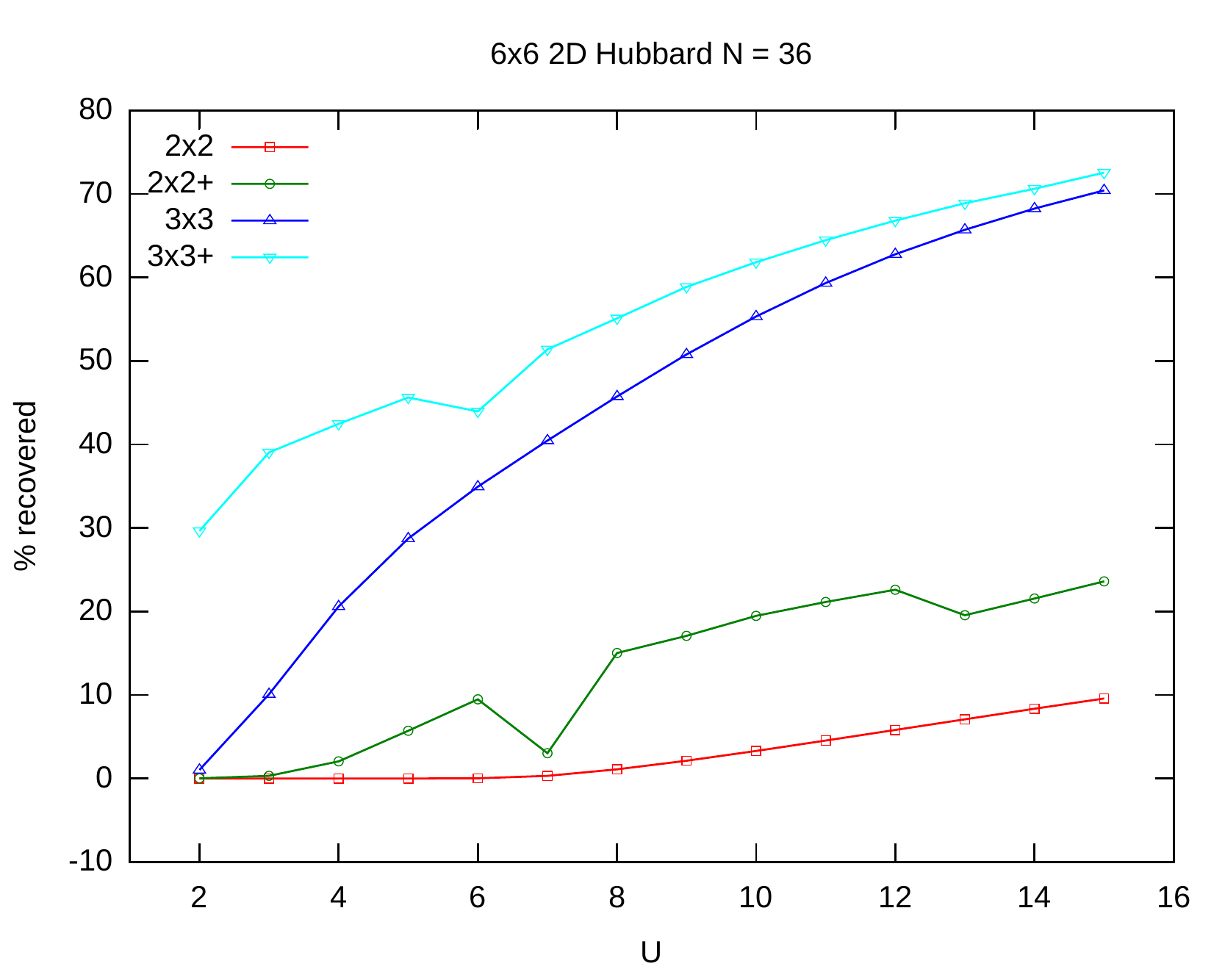}
\caption{\label{diff-36} Percentage of the gap between the $\mathcal{PQG}$ and the $\mathcal{PQGT}'$ results recovered by the addition of the pure and extended cluster constraints with optimized non-cluster states $p$ and $h$. For the $6\times6$ Hubbard model at half filling.}
\end{figure}

\begin{figure}
\centering
\includegraphics[scale=0.9]{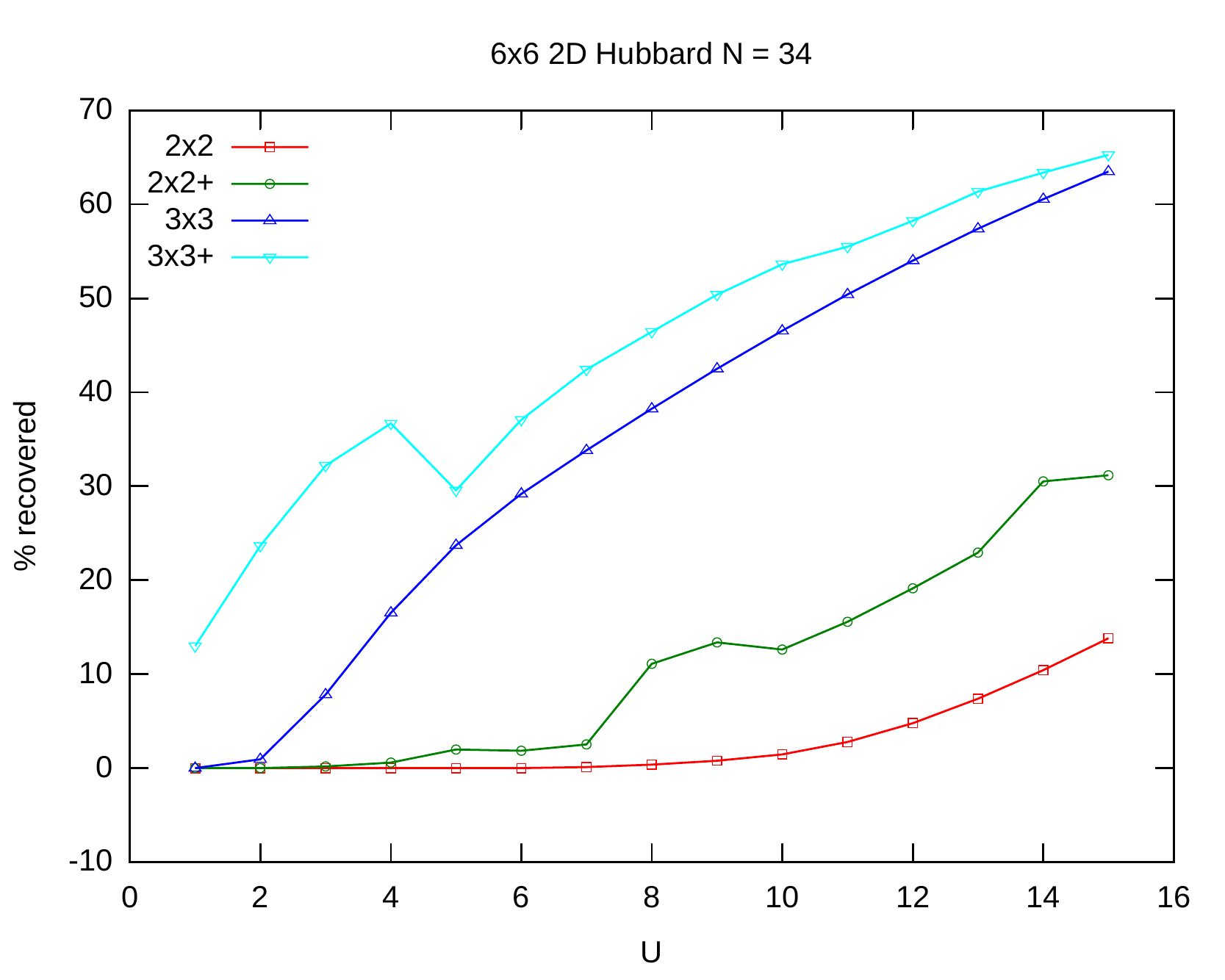}
\caption{\label{diff-34} Percentage of the gap between the $\mathcal{PQG}$ and the $\mathcal{PQGT}'$ results recovered by the addition of the pure and extended cluster constraints with optimized non-cluster states $p$ and $h$. For the $6\times6$ Hubbard model with 34 particles.}
\end{figure}

The procedure introduced in Section~\ref{optim_con} has been applied to optimize the $p$ state for $e\mathcal{T}_1$, and the $p$ and $h$ states for the $e\mathcal{T}_2'$ condition. In Figures~\ref{diff-36} and \ref{diff-34} we show the percentage of the gap between the results obtained by imposing only $\mathcal{PQG}$ constraints and the full $\mathcal{PQGT}'$ result that has been recovered by the addition of the pure cluster constraints, and the improvement caused by imposing the extended cluster constraints with optimized non-cluster states. For the $2\times2$ constraints one sees that, whereas the original cluster constraints have little to no effect, the extended constraints are active and improve the result. The extensions on the $3\times3$ cluster improve the result significantly for lower values of $U$, where the pure cluster constraints perform poorly. At higher values of $U$ they still enhance the result, although the pure cluster constraints already do a good job. We notice that the improvements are discontinuous as a function of $U$. This is probably a manifestation of the fact that the cost function is not convex, and the conjugate gradient algorithm ends up in different local optima for various $U$-values.

\section{Conclusion and outlook}
In this paper the ground-state energy of the 2D Hubbard model has been calculated using the v2DM method. Results have been obtained using the accurate $\mathcal{PQGT}'$ conditions for lattices up to $6\times6$ at different fillings. Hereby a rigorous variational \emph{lower bound} is obtained which can be used as a reference for future calculations in combination with the \emph{upper bound} obtained by the DMRG algorithm. Imposing the three-index constraints gives results that are of a significant better quality than imposing just the two-index conditions, but at a computational cost which prevents us from scaling to larger lattice sizes. We have therefore imposed the three-index conditions on local clusters of $2\times2$ and $3\times3$ sites, in an attempt to capture the relevant local correlation at the three-index level while avoiding the computational cost of the full $\mathcal{PQGT}'$ conditions. This worked reasonably well provided the cluster size was large enough. However, imposing these constraints is equivalent to treating the cluster as a closed system, whereas the cluster subsystem should rather be treated as an open system, including communication between the subsystem and the rest of the system. This observation led us to derive new constraints, which take the open-system nature into account, and are dependent on states outside of the cluster. We have shown how to choose these states by performing an eigenvalue optimization, which can be done efficiently using a conjugate gradient algorithm. The results of these proof-of-principle calculations show that adding a limited set of the extended cluster constraints already  improve the quality of the results substantially. The increase in accuracy was largest where the pure cluster constraints failed to recover a large part of the difference between $\mathcal{PQG}$ and $\mathcal{PQGT}'$. The improvements brought about by imposing these extended constraints come at a very small additional computational cost.

In this paper we have set forth the new conditions, and implemented a proof-of-principle example. There are, however, many improvements that can be made. A first point is that the constraint depends on our choice for the cost function. There is no guarantee that the constraint with the lowest eigenvalue will lead to the largest increase in energy when imposed. We have seen that local minima occur, because our cost function is non-linear, so the conjugate gradient algorithm is bound to get stuck there. Using a Monte Carlo optimization of the function, a global optimum of the cost function might be found. In this discussion we have limited ourselves to single states outside of the cluster. It can be expected, however, that multiple orthogonal states will provide better results. If we allowed two states outside the cluster, the $e\mathcal{T}_1$ would be constructed using the operator:
\begin{equation}
B^\dagger = \sum_{abc}t^1_{abc}a^\dagger_a a^\dagger_b a^\dagger_c+\sum_{ab}x^1_{ab}a^\dagger_a a^\dagger_b p^\dagger + \sum_{ab}x^2_{ab}a^\dagger_a a^\dagger_b p_{\perp}^\dagger~,
\end{equation}
which leads to another quadratic addition in the $\mathcal{T}_1$ matrix. If more and more terms of this sort are added one should converge to the full $\mathcal{PQGT}'$ result. Of course, the hope is that convergence is achieved by adding just a few states.

For the proof-of-principle calculations a limited set of the extended constraints have been used, namely those that have one-particle terms outside of the cluster. It is likely that correlations that include two-particle and particle-hole terms outside of the cluster are important, and that adding those would improve the result considerably.

As a final note we mention that these constraints are general and not limited to the 2D Hubbard model. They can be used in molecular calculations as an active space method to include three-index conditions on a limited space to recover a portion of the $\mathcal{PQGT}'$ result without the heavy computational burden.

\appendix
\section{\label{error_est}Error estimation of the DMRG results}
\begin{figure}
\begin{centering}
\includegraphics[scale=0.8]{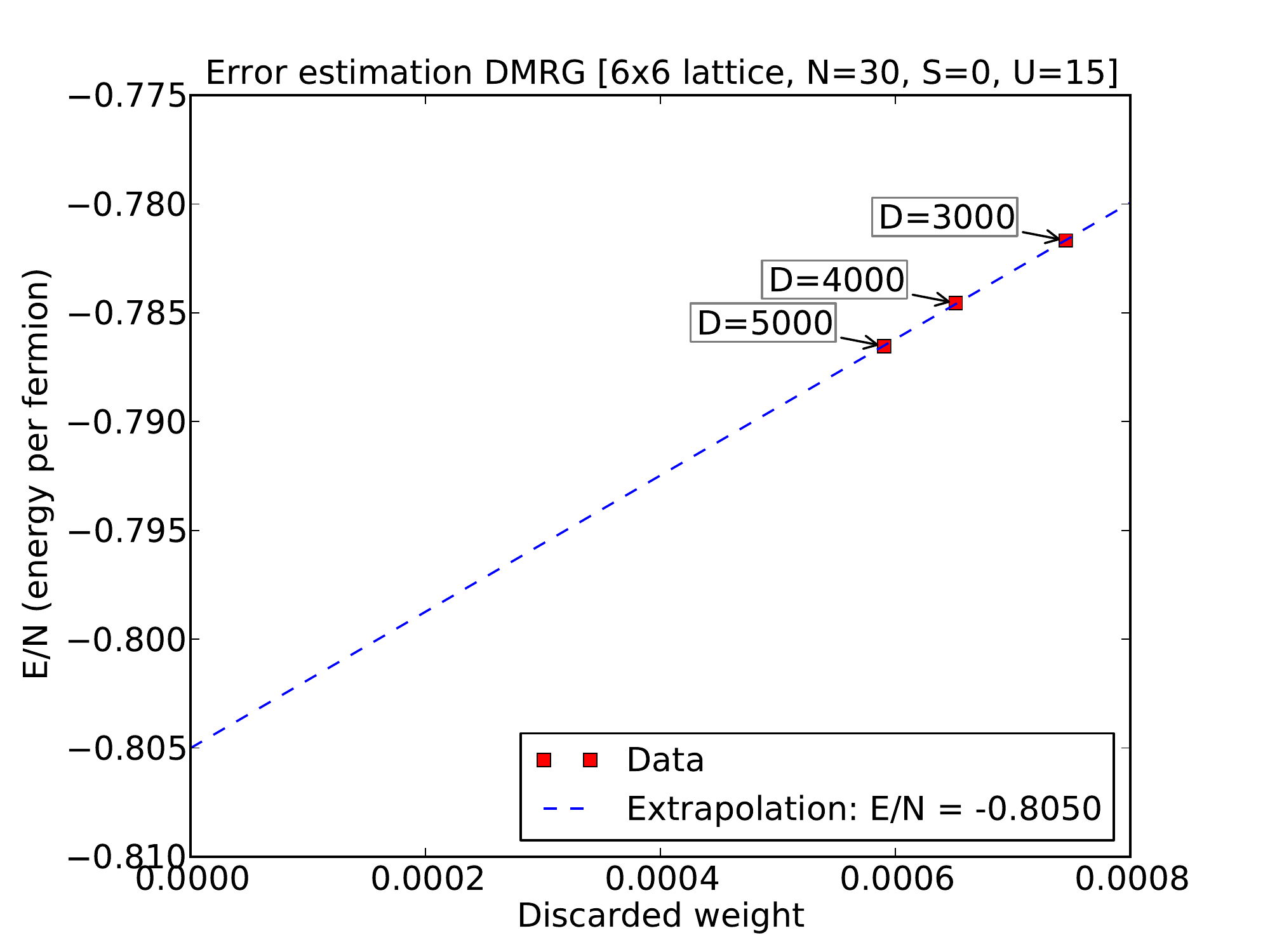}
\end{centering}
\caption{\label{DMRGextrapol} Rough estimate of the true ground state energy for the $6 \times 6$ lattice, filled with $N=30$ fermions, in the $S=0$ singlet state, with the on-site repulsion $U=15$. The estimate is based on a linear extrapolation between the discarded weight and the obtained variational energy, for several DMRG calculations using different values of reduced renormalized basis states $D$.}
\end{figure}

Underlying the DMRG algorithm is the class of matrix product state wave-functions. The minimal energy encountered during a DMRG sweep is therefore a variational upper bound to the exact ground state energy. This energy can be improved by increasing the number of retained renormalized basis states. To allow for an extrapolation to the true ground state energy, a linear relationship between the so-called discarded weight and the energy has been advocated \cite{legaza,chan_dmrg1,chan_dmrg2}. For several numbers of retained renormalized basis states, the maximal discarded weight and the minimal variational energy obtained from the last (converged) sweep are plotted. A linear fit then allows to obtain a rough estimate for the true ground state energy.\\

For the $6 \times 6$ lattice, filled with $N=30$ fermions, in the $S=0$ singlet state, with the on-site repulsion $U=15$, the v2DM and the DMRG results deviate the most in Table II. For this case, we have tried to obtain a rough estimate for the exact ground state energy in Fig. \ref{DMRGextrapol}, with the method described above. Note that $D$ is the number of \textit{reduced} renormalized basis states which are retained during the sweeps. The true ground state energy estimation for $N=30$ and $U=15$ allows to attribute $15\%$ of the corresponding v2DM-DMRG gap to DMRG and $85\%$ to v2DM.\\

\begin{acknowledgments}
We gratefully acknowledge financial support from FWO-Flanders and the research council of Ghent University. S.W. has a PhD fellowship from the FWO-Flanders. S.D.B and B.V. are post-doctoral fellows of the FWO-Flanders. The authors are Members of the QCMM alliance Ghent-Brussels.
\end{acknowledgments}
\bibliography{cluster.bib}
\end{document}